\documentclass[prd,aps,preprint,amsmath,nofootinbib,amssymb,eqsecnum]{revtex4-1}
\usepackage{verbatim,graphics,graphicx,color,slashed}
\usepackage{hyperref}
\newcommand{\GeV}{\textrm{ GeV}}
\newcommand{\TeV}{\textrm{ TeV}}

\begin{document}

\title{\Large Note on  125 GeV Spin-2 particle}
\author{ Chao-Qiang Geng${}^{a,b}$ }
\author{ Da Huang${}^{a}$ }
\author{ Yong Tang${}^{b}$ }
\author{ Yue-Liang Wu${^{c,d}}$}
\affiliation{${}^{a}$Department of Physics, National Tsing Hua University, Hsinchu, Taiwan
\\${}^{b}$Physics Division, National Center for Theoretical Sciences, Hsinchu, Taiwan
\\${}^{c}$Kavli Institute for Theoretical Physics China (KITPC), \\ State Key Laboratory of Theoretical Physics (SKLTP),\\
Institute of Theoretical Physics, Chinese Academy of Sciences, Beijing
\\${}^{d}$University of Chinese Academy of Sciences
}

\date{\today}

\begin{abstract}
A new boson around 125 GeV without specific spin
has been observed by both ATLAS and CMS at the LHC. Since its decay into a diphoton excludes the spin-1
case by the Landau-Yang theorem,it leaves 0 or 2 as the  possible lowest spin for the new boson.
Instead of the well-established spin-0 Higgs-like boson, we take this new boson to be a spin-2 massive Graviton-like particle
denoted as  $G$, which exists copiously in extra-dimension theories,  and concentrate on its phenomenology.
In particular, we calculate the three-body decays of  $G\rightarrow V f\bar{f}'$ with $V$ and $f^{(\prime)}$
the gauge boson and fermions in the standard model (SM) and compare our results with those of the SM Higgs boson.
The couplings between $G$ and $V$s are also estimated by fitting the data. A new observable that can distinguish $G$ from the Higgs is proposed.
\end{abstract}

\maketitle

\section{Introduction}
A new particle $H$ around 125 GeV has been observed by ATLAS \cite{atlas:2012gk} and CMS \cite{cms:2012gu}
with the combined significances of 5.9 and 5.0 standard deviations, respectively,
at the LHC. Excesses of events have been shown in various channels, such as
$H\rightarrow \gamma\gamma$, $H\rightarrow ZZ^{*}$ and $H\rightarrow WW^{*}$.
The next step is to have precision measurements
as well as determinations of the particle properties, such as its spin, $CP$ and decay branching ratios.

With the observation of $H\rightarrow \gamma\gamma$, the Landau-Yang theorem \cite{Landau, Yang}
implies that the spin of $H$ can not be 1. As this new particle must be a boson,
it leaves that  0 or 2 as the lowest possible spin for the particle.
Although at this moment the production and decay of this new particle are consistent with those of
the Higgs boson in the standard model (SM) within
$2\sigma$~\cite{Giardino:2012dp, Carmi:2012in, Ellis:2012hz, Espinosa:2012im, Klute:2012pu, Corbett:2012dm, Plehn:2012iz, Eberhardt:2012gv, Baak:2012kk, Moreau:2012da, Dawson:2012di},
the current data are not enough to tell its spin.
Since spin-0 bosons, such as Higgs, dilaton and radion, have been widely examined
in the literature~\cite{Barger:2011qn, deSandes:2011zs, Barger:2011hu, Cheung:2011nv,Tang:2012uw, Grzadkowski:2012ng, Davoudiasl:2012xd, Matsuzaki:2012vc, Low:2012rj, Elander:2012fk, Matsuzaki:2012xx, Chacko:2012vm, Bellazzini:2012vz},
in this note we shall concentrate on the spin-2 particle denoted as  $G$.

Spin-2 particles are copious in particle physics models, especially those with extra-dimensions. For example, two popular models, ADD~\cite{ArkaniHamed:1998rs} and Randall-Sundrum (RS)~\cite{Randall:1999ee} types, which were motivated to solve the hierarchy problem,
have a Kaluza-Klein (KK) tower for the spin-2 gravitons. Depending on the specific model, the couplings between the massive gravtion and SM particles could have various structures~\cite{Giudice:1998ck, Han:1998sg, Davoudiasl:2000wi}. Aiming at a wide application of our study,
we shall be only interested in the general framework rather than  a special model.

In this note, we shall present the relevant analytic formulas for the three-body decays of
$G\rightarrow VV^{*}$($V^{*}\rightarrow f\bar{f}$) with $V$ and $f$ the gauge boson and fermion in the SM.
 These three-body decays can be important when the mass of $G$, $m_G$, is smaller than $2m_V$. In addition, as discussed in Refs.~\cite{Ellis:2012xd, Alves:2012fb, Choi:2012yg, Choi:2002jk, Buszello:2002uu, Gao:2010qx, Bolognesi:2012mm, Kumar:2012ba,Ellis:2012wg, Englert:2012ct, Boughezal:2012tz, Stolarski:2012ps}, $ZZ^{*}\rightarrow 4l^{\pm}$ is the most relevant channel for a full determination of the spin of the new particle.
 Note that there are some computer codes, which may also produce numerical results~\cite{Hagiwara:2008jb, Alwall:2011uj} for the decays.

This paper is organized as follows. In Sec.~\ref{framework}, we establish the framework of our present study.
In Sec.~\ref{Compute}, we calculate the three-body decays of the spin-2 particle $G$ and present the full analytic formulas for the decay rates.
In Sec.~\ref{pheno}, we explore the phenomenology related to $G$.
Finally, a brief summary is given in Sec.~\ref{Sum}.

\section{Framework}\label{framework}
We start with the interactions, given by
\begin{equation}
\label{L}
\mathcal{L}_{\textrm{int}}=-\displaystyle\sum\limits_{i}\frac{c_i}{\Lambda}h_{\mu\nu}T^{\mu\nu}_{i},
\end{equation}
where $T^{\mu\nu}_{i}$ stand for the energy-momentum tensors of the SM particles, $h_{\mu\nu}$ is the field for the spin-2 particle,
 $c_i$ encode the relative coupling strengths between different particles, and
 $\Lambda$ characterizes the typical energy scale.

We remark that although the interacting terms in Eq.~(\ref{L})
are not the most general ones~\cite{Gao:2010qx, Bolognesi:2012mm}, they are typical for
many massive graviton or extra-dimensional models. The most general form can be given by including all the Lorentz and gauge invariant operators of dimension-5 at the lowest order. For simplicity, we shall only consider the interactions in Eq.~(\ref{L})
in the following discussions.

In models with all SM particles confined to 4 dimensions, $c_i$ are universal.
In such cases, the graviton is most likely to decay into $q\bar{q}$ and a gluon pair. Constraints on such models have been studied
 in Refs.~\cite{Tang:2012uw,Tang:2012pv} by using the latest dijet and dilepton searches at the LHC. The results show that if the mass of the graviton lies in the $\mathcal{O}(100 \GeV)$ region, $\Lambda$ is constrained stringently to be as large as $\mathcal{O}(10 \TeV)$.
However,
this constraint can be relaxed if we allow non-universal $c_i$. In general, $c_i$ may not be the same for different  particles~\cite{Davoudiasl:2000wi}.
For instance, in models in which  SM particles also propagate in extra-dimensions, $c_i$ depend on overlaps among the wave functions of KK modes for the interacting particles, so that $c_i$  could be different for SM particles.

\begin{figure}[t]
\includegraphics[scale=0.35]{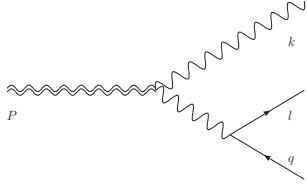}
\caption{Feynman diagram of order $\mathcal{O}(\alpha c^2_V)$ that dominantly
contributes to $G\rightarrow VV^*(V^*\rightarrow f\bar{f}')$, where the double line stands for the spin-2 particle,
  $P,k,l$ and $q$ represent  the momenta of $G, V, f$ and $\bar{f}'$,  and the directions of $k,l$ and $q$ are out of the vertex,
  respectively.}
\label{Fig:Dominant}
\end{figure}

\begin{figure}[t]
\includegraphics[scale=0.35]{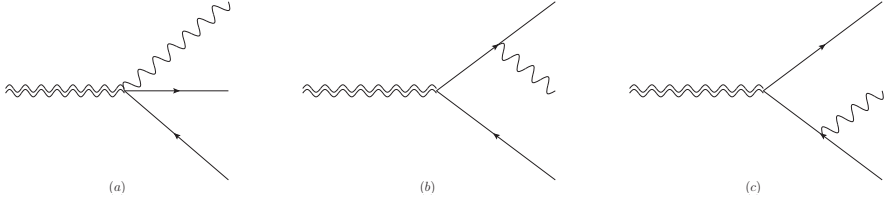}
\caption{Feynman diagrams of order $\mathcal{O}(\alpha c^2_f)$ that sub-dominantly
contribute to $G\rightarrow VV^{*}(V^*\rightarrow f\bar{f}')$ with small $c_f$.}
\label{Fig:Neglectable}
\end{figure}

In this paper, we  assume  small values of $c_i$ for the light fermions, $i.e.$ $c_f\ll 1$, and concentrate on
the phenomenology which is only  relevant to $c_{i}$ for the gauge bosons.
Consequently, for the decay processes of $G\rightarrow VV^*(V^*\rightarrow f\bar{f}')$, we only have to calculate the Feynman diagram in Fig.~\ref{Fig:Dominant},
which is of order $\mathcal{O}(\alpha c^2_V)$, whereas the diagrams in Fig.~\ref{Fig:Neglectable} of order $\mathcal{O}(\alpha c^2_f)$ can be
  neglected. In the following calculation, we further assume $c_W=c_Z$ and normalize $c_W=1$ by scaling $\Lambda$.

\section{Three-body decays of $G \to V  f \bar{f}^\prime$ }
\label{Compute}
We now compute the decays of $G \to V f \bar{f}^\prime$ ($V=W^{\pm}, Z$).
We are interested in the case in which the massive spin-2 particle $G$ lies in the range of $m_V < m_G < 2 m_V$.
It is clear that this massive particle is not kinematically allowed to decay into two on-shell gauge bosons.
 The only possibility is that it decays to a real gauge boson and a virtual one,
and the latter further  to two leptons or quarks.
Moreover, as discussed above, we  are assuming that $G$ couples to $W$ or $Z$ dominantly over light fermions. Otherwise, $G$ is more likely to decay into $f\bar{f}$ since two-body decay channels would be dominant. 
As a result, we can safely neglect the contributions from the direct couplings of $G$ and light fermion pairs, see Fig.~\ref{Fig:Neglectable}. We also ignore the effects of the fermion mass $m_f$ since it is much less than that of  the $V$-boson. Thus, the only relevant Feynman diagram for $ G\to V \bar{f} f^\prime$ is shown in Fig.~\ref{Fig:Dominant}.

For convenience, we give the Feynman rule for the three-point vertex of a massive spin-2 particle and two W-bosons \cite{Han:1998sg} with $\kappa=2/ \Lambda$ in the unitary gauge as follows:
\begin{eqnarray}
GWW &: & -i\frac{\kappa}{2}\{(M_W^2+k_1\cdot k_2) C_{\mu\nu,\rho\sigma} + D_{\mu\nu,\rho\sigma}(k_1,k_2)\},
\end{eqnarray}
where 
\begin{eqnarray}
C_{\mu\nu,\rho\sigma} &= & \eta_{\mu\rho} \eta_{\nu\sigma} +\eta_{\mu\sigma}\eta_{\nu\rho}-\eta_{\mu\nu}\eta_{\rho\sigma}, \nonumber\\
D_{\mu\nu,\rho\sigma}(k_1,k_2) &= &\eta_{\mu\nu}k_{1\sigma}k_{2\rho}-[\eta_{\mu\sigma} k_{1\nu} k_{2\rho} +\eta_{\mu\rho}k_{1\sigma}k_{2\nu}-\eta_{\rho\sigma}k_{1\mu}k_{2\nu}+ (\mu\leftrightarrow\nu)]. 
\end{eqnarray}
Here, the directions of two W-boson momenta $k_1$ and $k_2$ are defined to be flowing into the vertex. 
Other vertices are just the standard model ones. With this convention, the amplitude for $G\rightarrow W^{\pm}W^{*\mp}(W^{*\mp}\rightarrow f\bar{f}')$ 
is given by
\begin{equation}
\label{AM}
{\cal M} = \frac{i g \kappa}{2\sqrt{2}} \epsilon^s_{\mu\nu}(P)\epsilon^A_\rho (k) \Big{[}[m_W^2+(P-k)\cdot k]C_{\mu\nu,\rho\sigma}+D_{\mu\nu,\rho\sigma}(k,P-k)\Big{]}\frac{ \bar{u}(l)\gamma^\sigma P_L v (q)}{m_G^2-2 P\cdot k},
\end{equation}
where $P,k,l$ and $q$ denote the four momenta of $G, V, f$ and $\bar{f}'$, respectively.
We  note that the properties of polarization tensors of W-bosons $\epsilon^A_\mu(k)$ and spin-2 particles $\epsilon^s_{\mu\nu}(P)$ enforce
\begin{eqnarray}
 k^\mu \epsilon^A_\mu(k)=0, && \;\; \epsilon^{s,\mu}{}_\mu(P) =0, \nonumber\\
  P^\mu \epsilon^s_{\mu\nu}(P)=0, &&\;\;  \epsilon^{s,\mu\nu}(P) \epsilon^{s^\prime *}_{\mu\nu}(P) =\delta^{s s^\prime},
\end{eqnarray}
so that the terms proportional to $\eta_{\mu\nu}$, $k^\rho$, $P^\mu$ and $P^\nu$  vanish in the amplitude of Eq.~(\ref{AM}).
The completeness conditions for the W-boson and the massive spin-2 particle are 
\begin{eqnarray}
\sum^3_A \epsilon^A_\mu(k)\epsilon^{A*}_\nu = -(g_{\mu\nu}-\frac{k_\mu k_\nu}{M_W^2}),\nonumber\\
\sum^5_s \epsilon^s_{\mu\nu}(P) \epsilon^{s*}_{\rho\sigma}(P) = \frac{1}{2} B_{\mu\nu, \rho\sigma}(P),
\end{eqnarray}
respectively, where
\begin{eqnarray}
B_{\mu\nu,\rho\sigma}(P) &=&  (\eta_{\mu\rho}-\frac{P_\mu P_\rho}{M_G^2})(\eta_{\nu\sigma}-\frac{P_\nu P_\sigma}{M_G^2})+ (\eta_{\mu\sigma}-\frac{P_\mu P_\sigma}{M_G^2})(\eta_{\nu\rho}-\frac{P_\nu P_\rho}{M_G^2}) \nonumber\\
&&- \frac{2}{3}(\eta_{\mu\nu}-\frac{P_\mu P_\nu}{M_G^2})(\eta_{\rho\sigma}-\frac{P_\rho P_\sigma}{M_G^2}).
\end{eqnarray}
By squaring the amplitude, averaging over the five polarizations of the initial spin-2 particle, and summing over polarizations of the final W-boson,
we finally obtain
\begin{align}
\label{SAm}
\frac{1}{5}\sum_{s,A} |{\cal M}|^2 = \frac{g^2 \kappa^2}{5}
&\bigg{\{} \frac{4 m_W^2}{3 m_G^4} l\cdot P  q\cdot P (l\cdot P+q\cdot P)^2 +\frac{4l\cdot q}{3 m_G^4}(l\cdot P +q \cdot P)^2\Big{[}(l\cdot P)^2+(q\cdot P)^2\Big{]}\nonumber\\
&{} + \frac{10 m_W^4}{3m_G^2}l\cdot P  q\cdot P -\frac{m_W^2}{3m_G^2} l\cdot q \Big{[} 44 l\cdot P q\cdot P+9(l\cdot P)^2+9(q\cdot P)^2\Big{]}\nonumber\\
&{} + \frac{2 l\cdot q}{3 m_G^2}  \Big{[}l\cdot q ( l\cdot P )^2-5(l\cdot P)^2 q\cdot P - 5 l\cdot P (q\cdot P)^2+ 8\, l\cdot q l\cdot P  q\cdot P\nonumber\\
&{} + l\cdot q\,(q\cdot P)^2-5(l\cdot P)^3-5(q\cdot P)^3\Big{]}+\frac{10 m_G^2}{3}(l\cdot q)^2+ \frac{5 m_W^4}{3}  l\cdot q\nonumber\\
&{} -\frac{m_W^2  l\cdot q}{3}(13 l\cdot q-20l\cdot P-20 q\cdot P) +\frac{2}{3}  l\cdot q \Big{[}-15  l\cdot q (l\cdot P + q\cdot P )\nonumber\\
&{} + 11 (l\cdot q)^2+5(l\cdot P)^2+5(q\cdot P)^2\Big{]}\bigg{\}}.
\end{align}

For the spin-averaged amplitude, the general formula for the three-body decay rate is given by~\cite{Beringer:1900zz}
\begin{equation}
d \Gamma(G\to W f f^\prime) = \frac{1}{(2\pi)^3} \frac{1}{32 M_G^3} \frac{1}{5}\sum_{s,A} |{\cal M}|^2 dm_{12}^2 dm_{23}^2
\end{equation}
where $m_{12}^2= (l+q)^2$ and $m_{23}^2= (q+k)^2=(P-l)^2$. 
Consequently, we can rewrite Eq.~(\ref{SAm}) in terms of $m_{12}^2$ and $m_{23}^2$ by the following identities:
\begin{eqnarray}
l\cdot q = \frac{1}{2} m_{12}^2,\quad  l\cdot P=\frac{1}{2} (M_G^2-m_{23}^2), \quad  q\cdot P=\frac{1}{2}(m_{12}^2+m_{23}^2-M_W^2).
\end{eqnarray}
By integrating out the invariant mass $m_{23}^2$,  in the massive spin-2 particle rest frame we  derive the  partial decay rate to be 
\begin{eqnarray}
\label{dDW}
\frac{d\Gamma(G\to W f \bar{f}^\prime )}{dx} &=& \frac{g^2 m_G^3 \kappa^2}{92160\pi^3}\frac{(x^2-4\epsilon^2)^{1/2}}{(1-x)^2} [2(1-x) x^2(x^2-5x+10)\nonumber\\
&&+(3x^4+16x^3-46x^2+40)\epsilon^2 -8(3x^2+4x-14)\epsilon^4 +48 \epsilon^6 ],
\end{eqnarray}
where $x=2E_W/m_G$ and $\epsilon=m_W/m_G$ with
\begin{equation}
 2\epsilon \leq x\leq 1+\epsilon^2,\; m_W \simeq 80.4~\mbox{GeV},\; \frac{g^2}{4\pi}=\frac{\alpha}{\sin^2\theta_W}\simeq 0.034.
\end{equation}
 In the massive spin-2 particle rest frame, 
 Here, we have used
\begin{eqnarray}
m_{12}^2=(l+q)^2 = (P-k)^2 = M_W^2+M_G^2-2E_W M_G= M_G^2(1+\epsilon^2-x).
\end{eqnarray}

By integrating the variable  $x$ in Eq.~(\ref{dDW}), we get
the simple analytical formula for the decay rate of $G\to W f \bar{f}^\prime$, given by
\begin{eqnarray}\label{GWW}
 \Gamma(G\to W f \bar{f}^\prime )= \frac{g^2 m_G^3 \kappa^2}{92160\pi^3} F_G(\epsilon),
\end{eqnarray}
where
\begin{eqnarray}\label{GWX}
 F_G(\epsilon) &=& \frac{(368\epsilon^6+104\epsilon^4+29\epsilon^2-12)}{(4\epsilon^2-1)^{1/2}} \arccos \Big{[}\frac{3\epsilon^2-1}{2\epsilon^3}\Big{]} \nonumber\\
&&{} -\frac{1}{60}(21\epsilon^{10}-200\epsilon^8-9150\epsilon^6+4560\epsilon^4+2765\epsilon^2+2004)\nonumber\\
&&{} -(90\epsilon^6-30\epsilon^4+5\epsilon^2-12)\ln\epsilon.
\end{eqnarray}
This could be compared with the corresponding decay rate for the Higgs boson in the SM~\cite{Rizzo:1980gz, Keung:1984hn, Djouadi:2005gi}, given by
\begin{eqnarray}\label{HWW}
 \Gamma(H\to W f \bar{f}^\prime )= \frac{g^4 m_H }{3072\pi^3} F_H(\epsilon),
\end{eqnarray}
where
\begin{eqnarray}\label{HWX}
 F_H(\epsilon) &=& \frac{3(20\epsilon^4-8\epsilon^2+1)}{(4\epsilon^2-1)^{1/2}} \arccos \Big{[}\frac{3\epsilon^2-1}{2\epsilon^3}\Big{]} \nonumber\\
&&{} - \left(1-\epsilon^2 \right)\left(\frac{47}{2}\epsilon^{2}-\frac{13}{2}+\frac{1}{\epsilon^2}\right)-3\left(4\epsilon^4-6\epsilon^2+1 \right)\ln\epsilon.
\end{eqnarray}

Note that Eq.~(\ref{GWW}) only accounts for the single decay channel of the virtual $W$-boson.
However, experimentally it is more convenient to study the inclusive spin-2 particle decay of $G\to W^{\pm} X$,
for which we only need to multiply Eq.~(\ref{GWW}) by 18,\footnote{We only consider the distinct final states
with a $W^+$ or $W^-$ and light fermion pairs, excluding the top quark processes.}
 given by
\begin{eqnarray}
\Gamma(G\to W^{\pm}X)= \frac{g^2 m_G^3 \kappa^2}{5120\pi^3}\, F_G(\epsilon).
\end{eqnarray}

With the above analytical expression at hand, it is straightforward to write down the similar result for the case of $Z$:
\begin{eqnarray}\label{GZX}
 \Gamma(G\to Z X )= \frac{g^2 m_G^3 \kappa^2}{61440\pi^3\cos^2 \theta_W} (7-\frac{40}{3}\sin^2\theta_W +\frac{160}{9}\sin^4\theta_W)\, F_G(\epsilon^\prime),
\end{eqnarray}
where  $\epsilon^\prime=m_Z/m_G$ with
\begin{equation}
 m_Z\simeq 91.2\, \mbox{GeV} , \quad \sin^2\theta_W \simeq 0.231.
\end{equation}
The partial decay rates for the subprocess $G\to Z\bar{f}f$ can be easily obtained with multiplying Eq.~(\ref{GZX}) by each branching ratio, respectively.

\section{Phenomenology}\label{pheno}
We are now in a position to discuss the phenomenology of the spin-2 particle and compare our results with those of the Higgs boson in the SM.
The quantity we shall use is as follows:
\begin{equation}\label{eq:uxx}
\mu_{XX} = \frac{\sigma(pp\rightarrow G)\times \mathcal{B}r(G\rightarrow XX)}{\sigma(pp\rightarrow H_{\textrm{SM}})\times \mathcal{B}r(H_{\textrm{SM}}\rightarrow XX)},
\end{equation}
where $X$ stands for $W^\pm, Z$ and $\gamma$ for our interest. The main production channel is the gluon-gluon fusion due to the small value of $c_f$ and
the large parton distribution function (PDF) for gluons. Since in our framework the spin-2 particle can couple to gluons and photons at the tree level,
we  specify these couplings as $c_g$ and $c_\gamma$, respectively. Loop corrections are included by rescaling  $c_g$ and $c_\gamma$.

\begin{figure}[ht]
\begin{center}
\includegraphics{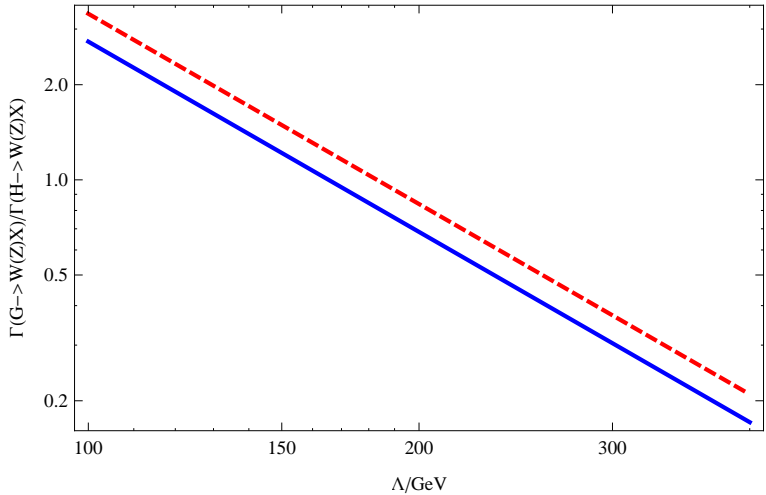}
\caption{Ratios of the spin-2 particle decay rates to those of the SM Higgs, where the dashed and solid lines correspond to
 $\Gamma(G\to W^{\pm}X)/\Gamma(H\to W^{\pm}X)$ and $\Gamma(G\to ZX)/\Gamma(H\to ZX)$, repectively.}\label{coupling}
\end{center}
\end{figure}

In order to estimate the scale $\Lambda$, it is useful to compare the spin-2 particle decay rates of
$\Gamma(G\to W^{\pm}X)$ and $\Gamma(G\to ZX)$ with the corresponding SM Higgs ones.
As shown in Fig.~\ref{coupling}, the spin-2 particle's decay rates, $\Gamma(G\to W(Z)X)$, are
 quite similar to those  of the Higgs  if $\Lambda$ is around $165$\GeV. We  note that $\Lambda$ and $c_g$ are correlated for fixed $\mu_{XX}$.
 For the convenience of further calculations, we  set  $\Lambda$ to be around $165\,\mbox{GeV}$ for $\Gamma(G\to ZX)/\Gamma(H_{\textrm{SM}}\to ZX)\sim 1$.

\begin{figure}[ht]
\begin{center}
\includegraphics{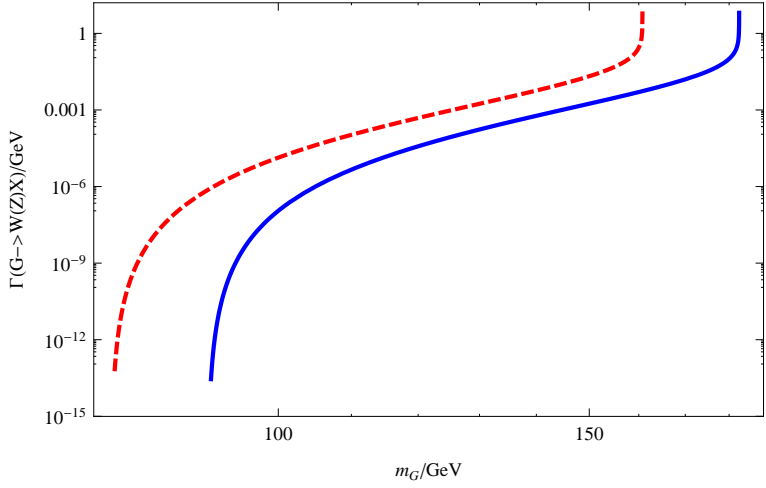}
\caption{Spin-2 particle decay rates as functions of  $m_G$, where the dashed and solid lines represent
 $\Gamma(G\to W^{\pm}X)$ and  $\Gamma(G\to ZX)$, respectively.}\label{mass}
\end{center}
\end{figure}

\begin{figure}[ht]
\begin{center}
\includegraphics{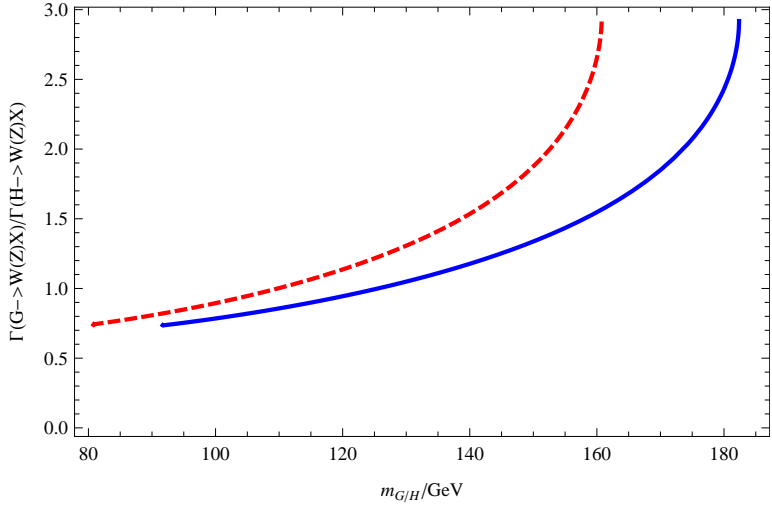}
\caption{Ratios of the spin-2 particle decay rates to those of the SM Higgs ones as  functions of
$m_G=m_H\equiv m_{G/H}$, where the dashed and solid lines stand for
$\Gamma(G\to W^{\pm}X)/\Gamma(H\to W^{\pm}X)$ and  $\Gamma(G\to ZX)/\Gamma(H\to ZX)$, respectively.}\label{mass2}
\end{center}
\end{figure}

For completeness, in Fig.~\ref{mass} we show the behavior of $G$ decaying to gauge bosons
as we vary the mass parameter of the spin-2 particle, $m_G$.
Note that in Fig.~\ref{mass} the two endpoints of  each line show singular behaviors.
This can be expected since our formula in Eq.~(\ref{GWX}) (Eq.~(\ref{GZX})) is only valid for the range of
$m_W<m_G<2 m_W$ ($m_Z<m_G<2 m_Z$), otherwise the particle would be either too light to produce a on-shell W (Z)
or heavy enough to decay into two on-shell Ws (Zs).
Furthermore, it would be helpful to compare our predicted spin-2 particle decay rates of these two weak interaction gauge boson channels with the corresponding SM Higgs ones for different mass parameters of $m_G$ and $m_H$ as
shown in Fig.~\ref{mass2}.

The behaviors of the curves in Figs.~\ref{mass} and \ref{mass2} are charactorised by $F_G$.
The difference between $F_G$ and $F_H$ can be an observable, simply because for $125\GeV$ we have
\[
9.10=\frac{\Gamma(G\rightarrow W^{\pm}X )}{\Gamma(G\rightarrow ZX )}\neq
 \frac{\Gamma(H_{\textrm{SM}}\rightarrow W^{\pm}X )}{\Gamma(H_{\textrm{SM}}\rightarrow ZX) }=8.14 \Rightarrow \frac{\mu_{WW}}{\mu_{ZZ}} = \frac{9.10}{8.14}.
\]
Consequently, if  the future  data shows that $\mu_{WW}/\mu_{ZZ}$ has some deviation from unity, this would imply new physical effects.

\begin{figure}[ht]
\begin{center}
\includegraphics[scale=0.3]{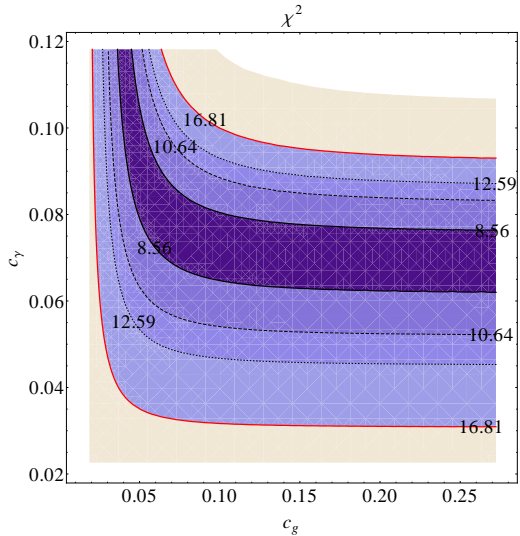}
\caption{$\chi^2$ fits as contours of $c_g$ and $c_\gamma$, where the darkest region around $c_\gamma \sim 0.08$ shows the best fit, while the
numerical numbers on the contours are the values of $\chi^2$.}
\label{Fig:Fit}
\end{center}
\end{figure}

Finally, we estimate the couplings $c_g$ and $c_\gamma$. Naively, if $\Gamma(G\rightarrow VV)\sim \Gamma(H_{\textrm{SM}}\rightarrow VV)$ ($V=g,\gamma$),
we obtain $c_g\sim 0.4$ and $c_\gamma\sim 0.2$, respectively. However, these naive estimations
just give rough values for $c_g$ and $c_\gamma$. Since $\mu_{WW}$ and $\mu_{ZZ}$ also depend on  $c_g$ and $c_\gamma$,
we shall perform a global fit to illustrate their possible allowed ranges.
We use the following data from ATLAS \cite{atlas:2012gk} and CMS \cite{cms:2012gu},
\begin{align*}
\textrm{ATLAS} &: \mu_{\gamma\gamma}=1.8\pm 0.5,\; \mu_{WW}=1.3\pm 0.5,\; \mu_{ZZ}=1.2\pm 0.6,\\
\textrm{CMS} &:   \mu_{\gamma\gamma}=1.56\pm 0.43,\; \mu_{WW}=0.6\pm 0.4,\; \mu_{ZZ}=0.7\pm 0.4.
\end{align*}
Our results are shown in Fig.~\ref{Fig:Fit}, where the darkest region corresponds to the best fitted parameters.
Explicitly, the region around $c_\gamma \sim 0.08$ and $c_g>0.1$ gives the best fit.
Note that  this region is insensitive to $c_g$ as the decay branching ratio is dominated by $G\rightarrow gg$, in which
the contribution from $c_g$ gets
almost cancelled in the numerator of $\mu_{XX}$ in Eq.~(\ref{eq:uxx}).

\section{Summary}
\label{Sum}
A new boson without specific spin around 125 GeV has been observed both by ATLAS and CMS  in the search for the Higgs boson in the SM at the LHC.
With the observation of  the diphoton mode, the Landau-Yang theorem excludes the particle to be a spin-1 particle.
The other possible lowest spin for the new particle is 0 or 2, corresponding to the Higgs-like or Graviton-like boson, respectively.
We have investigated the case that the new boson is identified as the spin-2 particle $G$.
We have calculated the rates of the three-body decay processes, $G\rightarrow Vf\bar{f}'$ ($V=W^\pm, Z$), and presented
the explicit analytic formulas with elementary functions, which as far as we know are not given previously in the literature.
These results could be useful for future studies.
 Phenomenology for $G$ has also been discussed. In particular,
 an observable of $\mu_{WW}/\mu_{ZZ}$, which can be used to distinguish between the SM Higgs and $G$, has been proposed.
 In addition, couplings between $G$ and gauge bosons have been estimated by fitting to the data.

\begin{acknowledgments}
The work by CQG, DH and YT was supported in part by National Center for Theoretical Science
and  National Science Council (NSC-98-2112-M-007-008-MY3 and
NSC-101-2112-M-007-006-MY3). The work by YLW was supported in part by
the National Science Foundation of China (NSFC) under Grant \#s of 10821504 and 10975170 and
the Project of Knowledge Innovation Program (PKIP) of the Chinese Academy of Science.
\end{acknowledgments}

\end{document}